\documentclass[
12pt, 
onesided, 
english, 
doublespacing, 
nolistspacing, 
liststotoc, 
headsepline, 
consistentlayout, 
openany,
]{MastersDoctoralThesis} 

\usepackage[utf8]{inputenc} 
\usepackage[T1]{fontenc} 

\usepackage{amsmath}
\usepackage{tabularx}

\usepackage{mathpazo} 
\usepackage{lipsum}
\usepackage{nameref}

\usepackage{varioref}
\usepackage{hyperref}
\usepackage{cleveref}

\usepackage{graphicx}
\usepackage{float}
\usepackage{cite}
\usepackage{epstopdf}
\usepackage{multirow}
\usepackage{amssymb}
\usepackage{amsmath}
\usepackage{amsthm}
\usepackage{mathtools, cases}
\usepackage{chngcntr}
\usepackage{amsmath,etoolbox}
\usepackage{amssymb}
\usepackage[ruled]{algorithm2e}
\usepackage{url}
\usepackage{caption}
\usepackage{subcaption}
\usepackage[font=small]{caption}
\usepackage{cleveref}
\usepackage{color}
\usepackage[utf8]{inputenc}

\newtheorem{theorem}{Theorem}
\newtheorem{lemma}{Lemma}
\makeatletter
\renewcommand\footnoterule{%
  \kern-3\p@
  \hrule\@width.4\columnwidth
  \kern2.6\p@}
  \makeatother

\newcommand{\E}{\mathbb{E}}

\newcommand{\setS}{\mathcal{S}}
\newcommand{\setC}{\mathcal{C}}

\thesistitle{Millimeter Wave Line-of-Sight Blockage Analysis} 
\supervisor{Shivendra Panwar} 
\examiner{} 
\degree{MASTER OF SCIENCE (Electrical Engineering)} 
\author{Ish Kumar Jain} 
\addresses{} 

\subject{Biological Sciences} 
\university{\href{http://engineering.nyu.edu/}{NEW YORK UNIVERSITY TANDON SCHOOL OF ENGINEERING}} 
\department{\href{http://engineering.nyu.edu/academics/departments/electrical}{ELECTRICAL AND COMPUTER ENGINEERING}} 
\group{\href{http://catt.nyu.edu/}{Center for Advanced Technology in Telecommunications (CATT)}} 
\faculty{\href{http://engineering.nyu.edu/people/shivendra-panwar}{Shivendra Panwar}} 

\AtBeginDocument{
\hypersetup{pdftitle=\ttitle} 
\hypersetup{pdfauthor=\authorname} 
\hypersetup{pdfkeywords=\keywordnames} 
}

\begin{document}

\frontmatter 

\pagestyle{plain} 


\begin{titlepage}
\begin{center}

\large  \textbf{\ttitle} \par 
\HRule \\ \vspace{0.2cm}

\large \textbf{THESIS} \par
\vspace{0.4cm}
 
\vfill

\large \textbf{Submitted in Partial Fulfillment of \\\vspace{0.2cm} the Requirements for \\\vspace{0.2cm} the Degree of \\\vspace{1cm} \degreename}\\[1cm]
at the\\[0.4cm]
\textbf{NEW YORK UNIVERSITY}\\
\textbf{TANDON SCHOOL OF ENGINEERING}\\[0.6cm]

by \\\vspace{1cm}

\textbf{\authorname} 
\vfill

\textbf{\large May 2018}\\[4cm] 

\vfill
\end{center}
\end{titlepage}
\begin{singlespacing}
\begin{center}

\large  \textbf{\ttitle} \par 
\HRule \\ \vspace{0.2cm}

\large \textbf{THESIS} \par
\vspace{0.4cm}
 

\large \textbf{Submitted in Partial Fulfillment of \\\vspace{0.2cm} the Requirements for \\\vspace{0.2cm} the Degree of \\\vspace{1cm} \degreename}\\[1cm]
at the\\[0.4cm]
\textbf{NEW YORK UNIVERSITY}\\
\textbf{TANDON SCHOOL OF ENGINEERING}\\[0.6cm]

by \\\vspace{1cm}

\textbf{\authorname} 
\\\vspace{1cm}
\textbf{\large May 2018}\\[0.5cm] 

\end{center}
\noindent
\hspace*{3.7in} Approved:\\
~\\
\hspace*{3.7in}\noindent\rule{2.3in}{0.4pt}\\
\hspace*{3.7in}{\footnotesize Advisor Signature}\\
~\\
\hspace*{3.7in}\noindent\rule{2.3in}{0.4pt}\\
\hspace*{3.7in}{\footnotesize Date}\\
 ~\\
\hspace*{3.7in}\noindent\rule{2.3in}{0.4pt}\\
\hspace*{3.7in}{\footnotesize Department Chair Signature}\\
~\\
\hspace*{3.7in}\noindent\rule{2.3in}{0.4pt}\\
\hspace*{3.7in}{\footnotesize Date}
\begin{tabbing}
University ID: \hspace{0.1in} \= \textbf{N11066411}\\
Net ID: \> \textbf{ikj211}
\end{tabbing}
\end{singlespacing}
\thispagestyle{empty}
\newpage
\begin{singlespacing}
\noindent Approved by the Guidance Committee:\\
~\\
~\\
\begin{tabbing}
\underline{Major:} \hspace{1in} \= Electrical and Computer Engineering\\
~\\
~\\
\hspace*{3in}\noindent\rule{3in}{0.4pt}\\
\hspace*{3in}\textbf{Shivendra S. Panwar}\\
\hspace*{3in}Professor\\
\hspace*{3in}Electrical and Computer Engineering\\
~\\
\hspace*{3in}\noindent\rule{2.2in}{0.4pt}\\
\hspace*{3in}Date\\
~\\
~\\
\hspace*{3in}\noindent\rule{3in}{0.4pt}\\
\hspace*{3in}\textbf{Elza Erkip}\\
\hspace*{3in}Institute Professor\\
\hspace*{3in}Electrical and Computer Engineering\\
~\\
\hspace*{3in}\noindent\rule{2.2in}{0.4pt}\\
\hspace*{3in}Date\\
~\\
~\\
\hspace*{3in}\noindent\rule{3in}{0.4pt}\\
\hspace*{3in}\textbf{Sundeep Rangan}\\
\hspace*{3in}Associate Professor\\
\hspace*{3in}Electrical and Computer Engineering\\
~\\
\hspace*{3in}\noindent\rule{2.2in}{0.4pt}\\
\hspace*{3in}Date\\
~\\
~\\
\end{tabbing}
 \end{singlespacing}
\newpage


\noindent{\LARGE Vitae}\vspace{0.2cm}\\
\indent Ish Kumar Jain was born in India. He received his Bachelor of Technology in Electrical Engineering from Indian Institute of Technology Kanpur, India, in May 2016. He received the Motorola gold medal for the best all-round performance in electrical engineering during his B.Tech.
\\
\indent Since 2016, he has been engaged in the Master of Science in Electrical Engineering at New York University,
Tandon School of Engineering in Brooklyn, New York. He was awarded the Samuel Morse MS fellowship to pursue research at NYU. He did an internship at Nokia Bell Labs during the summer of 2017 where he applied machine learning tools to future generation wireless communication systems.\\
\indent During his
M.S., he served as a Teaching Assistant for the Internet Architecture and Protocols lab in Spring 2017, and the Introduction to Machine Learning course in Fall 2017 and Spring 2018. His work on Millimeter-wave LOS blockage analysis has been accepted for publication as an invited paper at the International Teletraffic Congress (ITC) 2018~\cite{jain2018limited}. 
\bigbreak \newpage


\noindent{\LARGE Acknowledgments}\vspace{0.2cm}

First and foremost, I would like to thank my advisor Prof. Shivendra Panwar for
his guidance, inspiration, and constant support. He always gave me the freedom
to choose the direction of research I wanted to pursue and helped me to find the most interesting topic for my MS thesis. I am grateful for the long hours of discussion with him on my research work. His valuable pieces of advice helped me to grow as a person and a researcher.

 I am immensely grateful to Prof. Elza Erkip and Prof. Sundeep Rangan for their valuable guidance and support in my research projects. Many thanks for devoting their time to this thesis and serving on my committee. 
I am also thankful to Prof. Anna Choromanska who advised me on a Machine Learning side-project on proving the boosting-ability of tree-based multiclass classifiers. This work with Prof. Anna has been submitted for publication in the IEEE Transactions on Pattern Analysis and Machine Intelligence.
Finally, I thank Prof. Pei Liu, Prof. Yong Liu, and Prof. Yao Wang for their guidance through healthy discussions throughout my MS.

I was fortunate to have a perfect lab environment, which was only possible due
to my lab colleagues and many other friends.
I am particularly thankful to Rajeev Kumar for his active collaboration on my work.
This thesis would not have been possible without his advice
and the long discussions I had with him. 
I also thank Thanos Koutsaftis, Georgios Kyriakou, Nicolas Barati, Amir Hosseini, Fraida Fund, 
Shenghe Xu, Muhammad Affan Javed, Abbas Khalili, Amir Khalilian, Sourjya Dutta, George MacCartney, Chris Slezak, Kishore Suri, Prakhar Pandey, Arun Parthasarathy, and others, who made
working at NYU a pleasure. 
Gratitude should also go to Valerie Davis, Budget and Operations Manager for her constant support and paperwork throughout my
Masters. 

\begin{abstract}
\addchaptertocentry{\abstractname} 
Millimeter wave (mmWave) communication systems can provide high data rates but the system performance may degrade significantly due to mobile blockers and the user's own body. A high frequency of interruptions and long duration of blockage may degrade the quality of experience. For example, delays of more than about 10ms cause nausea to VR viewers. Macro-diversity of base stations (BSs) has been considered a promising solution where the user equipment (UE) can handover to other available BSs, if the current serving BS gets blocked.
However, an analytical model for the frequency and duration of dynamic blockage events in this setting is largely unknown.
In this thesis, we consider an open park-like scenario and obtain closed-form expressions for the blockage probability, expected frequency and duration of blockage events using stochastic geometry. Our results indicate that the minimum density of BS that is required to satisfy the Quality of Service (QoS) requirements of AR/VR and other low latency applications is largely driven by blockage events rather than capacity requirements. Placing the BS at a greater height reduces the likelihood of blockage. We present a closed-form expression for the BS density-height trade-off that can be used for network planning.

\end{abstract}


\tableofcontents 

\listoffigures 

\listoftables 


\begin{abbreviations}{ll} 
\textbf{LOS} & \textbf{L}ine \textbf{O}f \textbf{S}ight\\
\textbf{BS} & \textbf{B}ase \textbf{S}tation\\
\textbf{UE}&\textbf{U}ser \textbf{E}quipment\\
\textbf{PPP}&\textbf{P}oisson \textbf{P}oint \textbf{P}rocess\\
\textbf{QoS}&\textbf{Q}uality \textbf{o}f \textbf{S}ervice\\
\textbf{AR}&\textbf{A}ugmented \textbf{R}eality\\
\textbf{VR}&\textbf{V}irtual \textbf{R}eality\\
\textbf{CoMP}&\textbf{Co}ordinated \textbf{M}ulti-\textbf{P}oint\\
\textbf{RAN}&\textbf{R}adio \textbf{A}ccess\textbf{N}etwork\\
\textbf{BBU}&\textbf{B}ase\textbf{b}and \textbf{U}nit\\

\end{abbreviations}


\dedicatory{To my parents and Atul} 

\mainmatter 

\pagestyle{thesis} 

\chapter{Introduction}
\markboth{}{INTRODUCTION}
\section{Motivation}
Millimeter wave (mmWave) communication systems can provide high data rates of the order of a few Gbps~\cite{CellularCap-Rap}, suitable for the  Quality of Service (QoS) requirements for Augmented Reality (AR) and Virtual Reality (VR). For these applications, the user-equipment (UE) requires the data rate to be in the range of 100 Mbps to a few Gbps, and an end-to-end latency in the range of 1 ms to 10 ms~\cite{ATTARVR}. However, mmWave communication systems are quite vulnerable to blockages due to higher penetration losses and reduced diffraction~\cite{bai2015coverage}. Even the human body can reduce the signal strength by 20 dB~\cite{georgeFading}. Thus, an unblocked Line of Sight (LOS) link is highly desirable for mmWave systems. Furthermore, a mobile human blocker can block the LOS path between User Equipment (UE) and Base Station (BS) for approximately 500 ms~\cite{georgeFading}. The frequent blockages of mmWave LOS links and a high blockage duration can be detrimental to ultra-reliable and low latency communication (URLLC) applications. While many proposals to achieve low latency and high reliability have been proposed, such as  edge caching, edge computing, network slicing~\cite{7387263,6871674}, a shorter Transmission Time Interval (TTI), frame structure~\cite{7980747}, and flow queueing with dynamic sizing of the Radio Link Control (RLC) buffer at Data Link Layer~\cite{Kumar2018DynamicCO}, we focus our analysis on issues related to Radio Access Network (RAN) planning to achieve the stringent Quality of Service (QoS) requirements of URLLC applications.


One potential solution to blockages in the mmWave cellular network can be macro-diversity of BSs and coordinated multipoint (CoMP) techniques. These techniques have shown a significant reduction in interference and improvement in reliability, coverage, and capacity in the current Long Term Evolution-Advance (LTE-A) deployments and other communication networks~\cite{kim2011analysis, WILITV:Rkumar, Kumar2017WiLiTVAL}. Furthermore, Radio Access Networks (RANs) are moving towards the cloud-RAN architecture that implements macro-diversity and CoMP techniques by pooling a large number of BSs in a single centralized baseband unit (BBU)~\cite{IBMCloudRAN,chen2011c}. As a single centralized BBU handles multiple BSs, the handover and beam-steering time can be reduced significantly~\cite{cloudRAN}.  In order to provide seamless connectivity for ultra-reliable and ultra-low latency applications, the proposed 5G mmWave cellular architecture needs to consider key QoS parameters such as the probability of blockage events, the frequency of blockages, and the blockage duration. 
For instance, to satisfy the QoS requirement of mission-critical applications such as AR/VR, tactile Internet, and eHealth applications, 5G cellular networks target a service reliability of 99.999\% \cite{mohr20165g}.  
In general, the service interruption due to blockage events can be alleviated by caching the downlink contents at the BSs or the network edge~\cite{bastug2017toward}. However, caching the content more than about 10ms may degrade the user experience and may cause nausea to the users particularly for AR applications~\cite{westphal2017challenges}. 
An alternative when blocked is to offload traffic to sub-6GHz networks such as 4G, but this needs to be carefully engineered so as to not overload them.
Therefore, it is important to study the blockage probability, blockage frequency, and blockage duration to satisfy the desired QoS requirements. 

\section{Contribution}
This work presents a simplified blockage model for the LOS link using tools from the stochastic geometry. In particular, our contributions are as follow:
\begin{enumerate}
\item We provide an analytical model for dynamic blockage (UE blocked by mobile blockers) and self-blockage (UE blocked by the user's own body). The expression for the rate of blockage of LOS link is evaluated as a function of the blocker density, velocity, height and link length.

\item We evaluate the closed-form probability and expected frequency of simultaneous blockage of all BSs in the range of the UE. Further, we present an approximation for the expected duration of simultaneous blockage.  
\item We verify our analytical results through Monte-Carlo simulations by considering a random way-point mobility model for blockers.
\item Finally, we present a case study to find the minimum required BS density for specific mission-critical services and analyze the trade-off between BS height and density to satisfy the QoS requirements.
\end{enumerate}

\section{Related Work}
\label{sec:related-work}

A mmWave link may have three kinds of blockages, 
%
namely, static, dynamic, and self-blockage. Static blockage due to buildings and permanent structures has been studied in~\cite{bai2012using} and~\cite{bai2014analysis} using random shape theory and a stochastic geometry approach for urban microwave systems. The underlying static blockage model is incorporated into the cellular system coverage and rate analysis in\cite{bai2015coverage}. 
Static blockage may cause permanent blockage of the LOS link. However, for an open area such as a public park, static blockages play a small role.                                      
The second type of blockage is dynamic blockage due to mobile humans and vehicles (collectively called mobile blockers) which may cause frequent interruptions to the LOS link. Dynamic blockage has been given significant importance by 3GPP in TR 38.901 of Release 14~\cite{3gpptr}. An analytical model in~\cite{gapeyenko2016analysis} considers a single access point, a stationary user, and blockers located randomly in an area. The model in~\cite{wang2017blockage} is developed for a specific scenario of a road intersection using a Manhattan Poisson point process model.
MacCartney \textit{et al.}~\cite{georgeFading} developed a Markov model for blockage events based on measurements on a single BS-UE link. Similarly, Raghavan \textit{et al.}~\cite{raghavan2018statistical} fits the blockage measurements with various statistical models. However, a model based on experimental analysis is very specific to the measurement scenario. The authors in~\cite{han20173d} considered a 3D blockage model and analyzed the blocker arrival probability for a single BS-UE pair. 
Studies of spatial correlation and temporal variation in blockage events for a single BS-UE link are presented in~\cite{samuylov2016characterizing} and~\cite{gapeyenko2017temporal}. However, their analytical model is not easily scalable to multiple BSs, important when considering the impact of macro-diversity..

Apart from static and dynamic blockage, self-blockage plays a key role in mmWave performance.
The authors of~\cite{abouelseoud2013effect} studied human body blockage through simulation. A statistical self-blockage model is developed in~\cite{raghavan2018statistical} through experiments considering various modes (landscape or portrait) of hand-held devices. The impact of self-blockage on received signal strength is studied in~\cite{bai2014analysis} through a stochastic geometry model. They assume the self-blockage due to a user's body blocks the BSs in an area represented by a cone.

All the above blockage models consider the UE's association with a single BS. 
Macro-diversity of BSs is considered as a potential solution to alleviate the effect of blockage events in a cellular network.
The authors of~\cite{zhu2009leveraging} and~\cite{zhang2012improving} proposed an architecture for macro-diversity with multiple BSs and showed the  improvement in network throughput. A blockage model with macro-diversity is developed in~\cite{choi2014macro} for independent blocking and in~\cite{gupta2018macrodiversity} for correlated blocking. However, they consider only static blockage due to buildings.

The primary purpose of the blockage models in previous papers was to study the 
coverage and capacity analysis of the mmWave system. However, 
apart from the signal degradation, blockage frequency and duration also affects the performance of the mmWave system and are critically important for applications such as AR/VR.   
In this paper, we present a simple closed-form expression for a compact analysis to provide insight into the optimal density, height and other design parameters and trade-offs of BS deployment.

\section{Organization}
The rest of the thesis is organized as follows. The system model is described in Chapter \ref{ch01_system_model}. A generalized blockage model considering dynamic and self-blockages is presented in Chapter \ref{ch02_generalized}. Chapter \ref{ch03_blockage_events} generalizes the blockage model to multiple BSs and evaluates the key blockage metrics---blockage probability, blockage frequency, and the blockage duration. We present simulation results and verify our theoretical formulations in Chapter \ref{ch04_results}. Finally, we conclude this thesis and discuss the future work in Chapter \ref{ch.conc}.


\chapter{System Model}\label{ch01_system_model}
\chaptermark{System Model}

\section{BS model and blockage model}
\label{sec:system-model}
Our system model consists of the following settings:
\begin{itemize}
\item \textit{BS Model}: The mmWave BS locations are modeled as a homogeneous Poisson Point Process (PPP) with density $\lambda_T$. 
Consider a disc $B(o,R)$ of radius $R$ and centered around the origin $o$, where a typical UE is located.
We assume that each BS in $B(o,R)$ is a potential serving BS for the UE. 
Thus, the number of BSs $M$ in the disc $B(o,R)$ of area $\pi R^2$ follows a Poisson distribution with parameter $\lambda_T\pi R^2$, \textit{i.e.},
\begin{equation}\label{eqn:poisson}
    P_M(m) = \frac{[\lambda_{T} \pi R^2]^m}{m!}e^{-\lambda_{T} \pi R^2}.
\end{equation}  


Given the number of BSs $m$ in the disc $B(o,R)$, we have a uniform probability distribution for BS locations.
The BSs distances $\{R_i\}$ $\forall i=1,\ldots,m$ from the UE are independent and identically distributed (iid) with distribution
\begin{equation}\label{eqn:distribution}
 f_{R_i|M}(r|m) = \frac{2r}{R^2}; \ 0< r\le R, \forall\ i=1,\ldots,m.
\end{equation}
\begin{figure}[!t]
	\centering
	\includegraphics[width=0.85\textwidth]{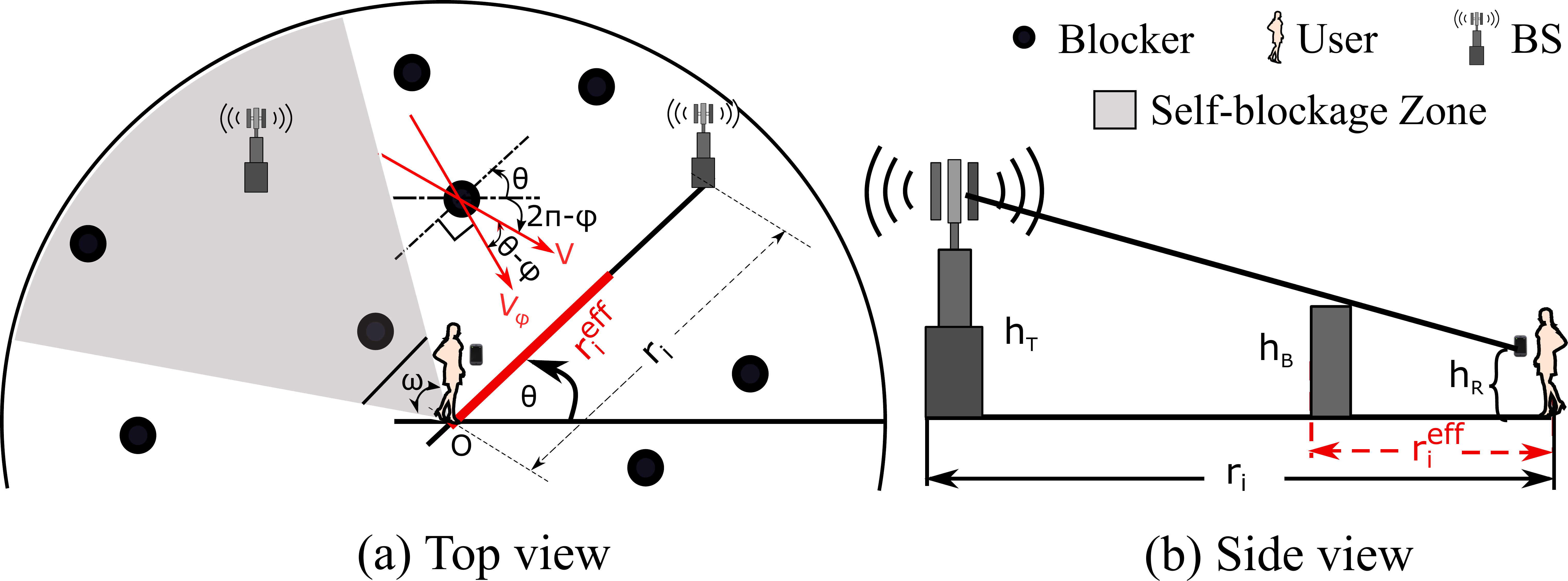}
	\caption{System Model}
	\label{fig:sysMod}
	\vspace{-4mm}
\end{figure}


\item \textit{Self-blockage Model}: The user blocks a fraction of BSs due to his/her own body. The self-blockage zone is defined as a sector of the disc $B(o,R)$ making an angle $\omega$ towards the user's body as shown in Figure~\ref{fig:sysMod}(a). Thus, all of the BSs in the self-blockage zone are considered blocked. 

\item \textit{Dynamic Blockage Model}: 
The blockers are distributed according to a homogeneous PPP with parameter $\lambda_{B}$. Further, the arrival process of the blockers crossing the $i^{th}$ BS-UE link is Poisson with intensity $\alpha_i$. The blockage duration is independent of the blocker arrival process and is exponentially distributed with parameter $\mu$. 



\item \textit{Connectivity Model}: We say the UE is blocked when all of the potential serving BSs in the disc $B(o,R)$ are blocked simultaneously.

\end{itemize}

\section{Single BS-UE link}
For a sound understanding of the system model, consider a single BS-UE LOS link in Figure~\ref{fig:sysMod}(a). The distance between the $i^{th}$ BS and the UE is $r_i$ and the LOS link makes an angle $\theta$ with respect to the positive x-axis. Further, the blockers in the region move with constant velocity $V$ at an angle $\varphi$ with the positive x-axis, where $\varphi\sim \text{Unif}([0,2\pi])$. 
Note that only a fraction of blockers crossing the BS-UE link will be blocking the LOS path, as shown in Figure \ref{fig:sysMod}(b). The effective link length $r_i^{eff}$ that is affected by the blocker's movement is 
\begin{equation}
	r_i^{eff}=\frac{\left(h_B-h_R\right)}{\left(h_T-h_R\right)}r_i,
\end{equation}
where $h_B, h_R$, and $h_T$ are the heights of blocker, UE (receiver), and BS (transmitter) respectively. The blocker arrival rate $\alpha_i$ (also called the blockage rate) is evaluated in Lemma \ref{lemma:alphan}. 

\begin{lemma}\label{lemma:alphan}
 The blockage rate $\alpha_i$ of the $i^{th}$ BS-UE link is
\begin{equation}\label{eqn:SingleBS}
\alpha_{i}
=Cr_i,
\end{equation}
where $C$ is proportional to blocker density $\lambda_B$ as,
\begin{equation}\label{eqn:C}
C = \frac{2}{\pi}\lambda_{B} V\frac{(h_B-h_R)}{(h_T-h_R)}.
\end{equation} 

\end{lemma}
\begin{proof}
Consider a blocker moving at an angle $(\theta-\varphi)$ relative to the BS-UE link (See Figure \ref{fig:sysMod}(a)). The component of the blocker’s velocity perpendicular to the BS-UE link is $V_\varphi = V\sin(\theta-\varphi)$, where $V_\varphi$ is positive when $(\theta-\pi)<\varphi<\theta$. 
Next, we consider a rectangle of length $r_i^{eff}$ and width $V_\varphi\Delta t$. The blockers in this area will block the LOS link over the interval of time $\Delta t$. Note there is an equivalent area on the other side of the link. Therefore, the frequency of blockage is $2\lambda_{B}r_i^{eff}V_\varphi\Delta t = 2\lambda_{B} r_i^{eff}V\sin(\theta-\varphi)\Delta t$. Thus, the frequency of blockage per unit time is $2\lambda_{B} r_i^{eff}V\sin(\theta-\varphi)$. Taking an average over the uniform distribution of $\varphi$ (uniform over $[0,2\pi]$), we get the blockage rate $\alpha_i$ as
\begin{equation}\label{eqn:SingleBS}
\begin{split}
\MoveEqLeft\alpha_{i} = 2\lambda_{B} r_i^{eff}V\int_{\varphi = \theta-\pi}^{\theta}\sin(\theta-\varphi)\frac{1}{2\pi}\,d\phi \\
\MoveEqLeft \qquad \quad =\frac{2}{\pi}\lambda_{B} r_i^{eff}V = \frac{2}{\pi}\lambda_{B} V\frac{(h_B-h_R)}{(h_T-h_R)}r_i.
\end{split}
\end{equation}
This concludes the proof. 
\end{proof}
Following~\cite{gapeyenko2017temporal}, we model the blocker arrival process as Poisson with parameter $\alpha_i$ blockers/sec (bl/s).
Note that there can be more than one blocker simultaneously blocking the LOS link. The overall blocking process has been modeled in~\cite{gapeyenko2017temporal} as an alternating renewal process with alternate periods of blocked/unblocked intervals, where the distribution of the blocked interval is obtained as the busy period distribution of a general $M/GI/\infty$ system.
For mathematical simplicity, we assume the blockage duration of a single blocker is exponentially distributed with parameter $\mu$, thus, forming an $M/M/\infty$ queuing system. We further approximate the overall blockage process as an alternating renewal process with exponentially distributed periods of blocked and unblocked intervals with parameters $\alpha_i$ and $\mu$ respectively. This approximation works for a wide range of blocker densities as shown in Section \ref{sec:numResults}. This approximation is also justified in Lemma \ref{lemma:Ps} as follow

\begin{lemma}\label{lemma:Ps}
Let $P_S$ denote the probability
that there are two or more blockers simultaneously blocking the single BS-UE link. 
Then, in order to have $P_S\le\epsilon$, the blockage rate $\alpha_i$ satisfies
\begin{equation}
	1 - e^{\alpha_i/\mu}\left(1+\alpha_i/\mu\right) \leq \epsilon,
\end{equation}
where $\alpha_i$ in (\ref{eqn:SingleBS}) is proportional to the blocker density $\lambda_B$. 
\end{lemma}
\begin{proof}
The probability $\mathbb{P}[\mathcal{S}_j]$ of event $\mathcal{S}_j$ is calculated as the $j^{th}$ state probability of the $M/M/\infty$ system. Therefore, we have
\begin{equation}
\begin{split}
P_S &= \mathbb{P}[\mathcal{S}_2] + \mathbb{P}[\mathcal{S}_3] + \cdots  + \mathbb{P}[\mathcal{S}_N] \\
	&= 1- \mathbb{P}[\mathcal{S}_0] - \mathbb{P}[\mathcal{S}_1]\\
    &= 1- e^{\alpha_i/\mu} -\frac{\alpha_i}{\mu}e^{\alpha_i/\mu} \\
    &=1 - e^{\alpha_i/\mu}\left(1+\alpha_i/\mu\right),
\end{split}
\end{equation}
Hence, for $P_S\le\epsilon$, we have $1 - e^{\alpha_i/\mu}\left(1+\alpha_i/\mu\right)\le\epsilon$
\end{proof}

\begin{figure}[!t]
	\centering
	\includegraphics[width=0.9\textwidth]{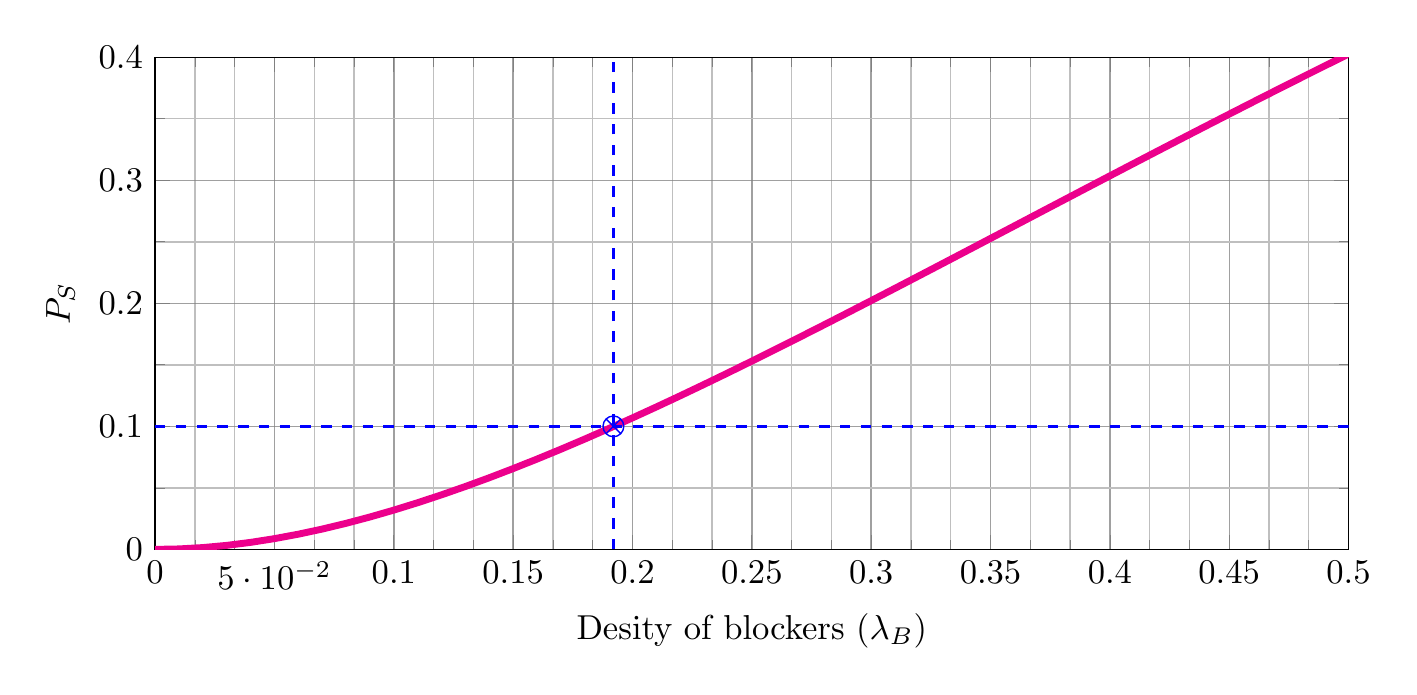}
	\caption{Probability that there are two or more blockers simultaneously blocking a single BS-UE link. The distance between BS and UE is 100 meters.}
	\label{fig:blockproab}
	\vspace{-6mm}
\end{figure}
From Lemma \ref{lemma:Ps}, we can say that the probability $P_S$ is low for lower blocker density $\lambda_B$ as shown in Figure~\ref{fig:blockproab}. In the worst case scenario when the distance between BS and UE is 100 m, i.e., $r_n=100$, in order to have $P_S \leq 0.1$, the blocker density $\lambda_B$ can be as high as 0.2 bl/m$^2$ and the blocker arrival rate $\alpha_i$ can be as high as 1.1 bl/sec. In fact, based on the real measurements obtained in Brooklyn, New York, USA, George \textit{et. al.,} have shown that the blockage rate of a single BS-UE link is approximately 0.2 bl/s in an urban scenario~\cite{georgeFading}. Therefore, as a first-order approximation and for reasonable values of blocker densities, we can ignore the events of simultaneously having more than one blockers in the blockage zone. 





In the next chapter, we consider a generalized blockage model for $M$ BSs which are in the range of UE. The UE keeps track of all these $M$ BSs using well-designed beam-tracking and handover techniques. 
Since the handover process is assumed to be very fast, the UE can instantaneously connect to any unblocked BS when the current serving BS gets blocked. Therefore, we consider the total blockage event occurs only when all the potential serving BSs are blocked. 

 \chapter{Generalized Blockage Model}\label{ch02_generalized}
\chaptermark{Generalized Model}

\section{Coverage Probability under Self-blockage}
Let there are $N$ BSs out of total $M$ BSs within the range of the UE that are not blocked by self-blockage. 
\begin{lemma}\label{lemma:PN}
The distribution of the number of BSs ($N$) outside the self-blockage zone and in the disc $B(o,R)$ is
\begin{equation}\label{eqn:PN}
P_N(n) = \frac{[p\lambda_{T} \pi R^2]^n}{n!}e^{-p\lambda_{T} \pi R^2},
\end{equation}
where \begin{equation}\label{eqn:p}
p= 1-\omega/2\pi 
\end{equation} 
is the probability that a randomly chosen BS lies outside the self-blockage zone in the disc $B(o,R)$. 
\end{lemma}
\begin{proof}
Due to the uniformity of BSs locations in $B(o,R)$, the distribution of $N$ given $M$ follows a binomial distribution with parameter $p=1-\frac{\omega}{2\pi}$, \textit{i.e.},
\begin{equation}\label{eqn:Bl_self}
P_{N|M}(n|m) = \binom{m}{n} (p)^{n}(1-p)^{m-n},\quad n\le m.
\end{equation}
The marginal distribution of $N$ is obtained as
\begin{equation*}
\begin{split}
P_N(n) &= \sum_{m=n}^\infty P_{N|M}(n|m)P_M(m)\\
&=\sum_{m=n}^\infty \binom{m}{n} (p)^{n}(1-p)^{m-n} \frac{[\lambda_{T} \pi R^2]^m}{m!}e^{-\lambda_{T} \pi R^2}\\
&= \frac{[p\lambda_{T} \pi R^2]^n}{n!}e^{-p\lambda_{T} \pi R^2} \sum_{m=n}^\infty \frac{1}{(m-n)!}[(1-p)\lambda_T \pi R^2]^{m-n}e^{-[(1-p)\lambda_T \pi R^2]}
\\
&=\frac{[p\lambda_{T} \pi R^2]^n}{n!}e^{-p\lambda_{T} \pi R^2}.
\end{split}
\end{equation*}
This concludes the proof.
\end{proof}
Let $\mathcal{C}$ denotes an event that the UE has at least one serving BS in the disc $B(o,R)$ and outside $S(o,R,\omega)$, \textit{i.e.}, $N\ne 0$. The probability of the event $\mathcal{C}$ is called the coverage probability under self-blockage and calculated as,
\begin{equation}\label{eqn:coveragep}
P(\mathcal{C}) =1- e^{-p\lambda_T\pi R^2}.
\end{equation}
The proof follows directly from (\ref{eqn:PN}) by putting $n=0$.

\section{Generalized Model with Dynamic Blockage}
\label{sec:generalized}

Given there are $n$ BSs in the communication range of UE and are not blocked by user's body, they can still get blocked by mobile blockers. The blocking event of these $N$ BSs is assumed to be independent.
Our objective is to develop a blockage model for the mmWave cellular system where the UE can connect to any of the potential serving BSs. In such a system, the blockage probability, expected blockage frequency, and expected blockage duration are the defining QoS parameters for low and ultra-low latency applications such as AR/VR.  
Since the blockage events form an alternating renewal process with exponentially distributed duration of blocked and unblocked intervals, we formulate a Markov chain to obtain the probability and duration of the simultaneous blockage of $k$ BSs out of the total $N$ available BSs. 

\begin{figure}[!t]
    \centering    
    	\includegraphics[width=0.7\textwidth]{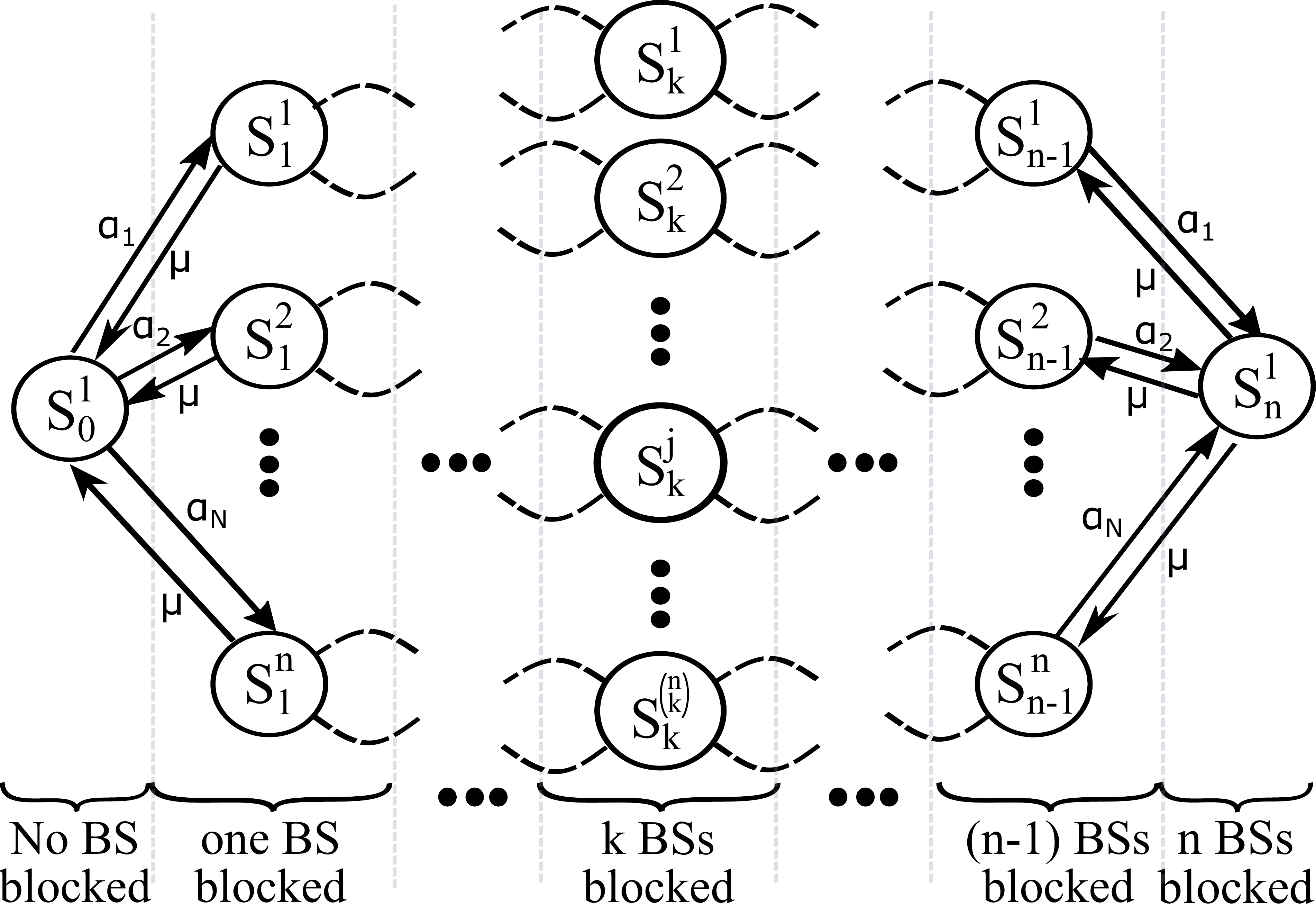}
    	\caption{Markov Chain for $N$ BSs}
    	\label{fig:MarkovNcase}
\vspace{-6mm}
\end{figure}
The Markov chain shown in  Figure~\ref{fig:MarkovNcase} has $2^n$ states where each state represents the blockage of a subset of n BSs. Let a set $\setS = \{1,2,\cdots,n\}$ represents the $n$ BSs. Define $\mathcal{P}(\setS)$ as a power set of $\setS$ of size $2^n$. The elements of $\mathcal{P}$ can be represented by a set $\setS_k^j$ for $j=1,\cdots,\binom{n}{k}$ and $k=0,\cdots,n$. Note that the set $\setS_k^j$ is associated with the state $S_k^j$ of our Markov model  shown in  Figure~\ref{fig:MarkovNcase}. Also, it is clear that the total number of states are $\sum_{k=0}^n\binom{n}{k}=2^n$. We are interested in the probability of the last state when $k=n$, since it represents the probability that all $n$ available BSs are blocked.







\begin{lemma}
Let $P_n^1$ denote the probability of the last state ($S_n^1$) of the Markov model in Figure \ref{fig:MarkovNcase}, then 
\begin{equation}\label{eqn:prob_stateN}
P_n^1  = \prod_{i=1}^n\frac{\alpha_i/\mu}{1+\alpha_i/\mu}  =\prod_{i=1}^n\frac{(C/\mu)r_i}{1+(C/\mu)r_i},
\end{equation}
where $C$ is defined in (\ref{eqn:C}).
\end{lemma}
\begin{proof}
The equilibrium steady-state distribution exists when $\alpha_i<\mu,\;\forall i=1:n$. These steady-state probabilities are derived as follow

The state probabilities of our Markov model in Figure~\ref{fig:MarkovNcase} are computed as 
\begin{equation}
P_k^j = \prod_{l\in\setS_k^j}\frac{\alpha_l}{\mu}P_0^1, \quad j=1,\cdots,\binom{n}{k},\quad k=1,\cdots,n,
\end{equation}
where $P_0^1$ represents the probability of the state $\setS_0^1$. By putting the sum of all state probabilities to 1, We obtain 
\begin{equation}
P_0^1 = \frac{1}{1+\sum_{k=1}^n\sum_{j=1}^{\binom{n}{k}}\prod_{l\in\setS_k^j}\frac{\alpha_l}{\mu}} = \frac{1}{\prod_{i=1}^n(1+\alpha_i/\mu)}.
\end{equation}
Therefore, the state probabilities become
\begin{equation}
P_k^j = \frac{\prod_{l\in\setS_k^j}\frac{\alpha_l}{\mu}}{\prod_{i=1}^n(1+\alpha_i/\mu)}, \quad j=1,\cdots,\binom{n}{k},\quad k=1,\cdots,n,
\end{equation}
Note the sum of state probabilities for $j=1,\cdots,\binom{n}{k}$ and for fixed $k$ represents the probability $P_k$ of the blockage of $k$ BSs, \textit{i.e.},
\begin{equation}
P_k = \sum_{j=1}^{\binom{n}{k}}P_k^j =  \frac{\sum_{j=1}^{\binom{n}{k}}\prod_{l\in\setS_k^j}\frac{\alpha_l}{\mu}}{\prod_{i=1}^n(1+\alpha_i/\mu)},\quad k=1,\cdots,n,
\end{equation}
By putting $k=n$, we get the probability that all $n$ BSs are blocked
\begin{equation}
P_n^1 =  \frac{\sum_{j=1}^{\binom{n}{n}}\prod_{l=1}^n\frac{\alpha_l}{\mu}}{\prod_{i=1}^n(1+\alpha_i/\mu)} = \prod_{i=1}^n\frac{\alpha_i/\mu}{1+\alpha_i/\mu}.
\end{equation}
\end{proof}

\begin{figure}[!t]
    \centering
   	\includegraphics[width=0.7\textwidth]{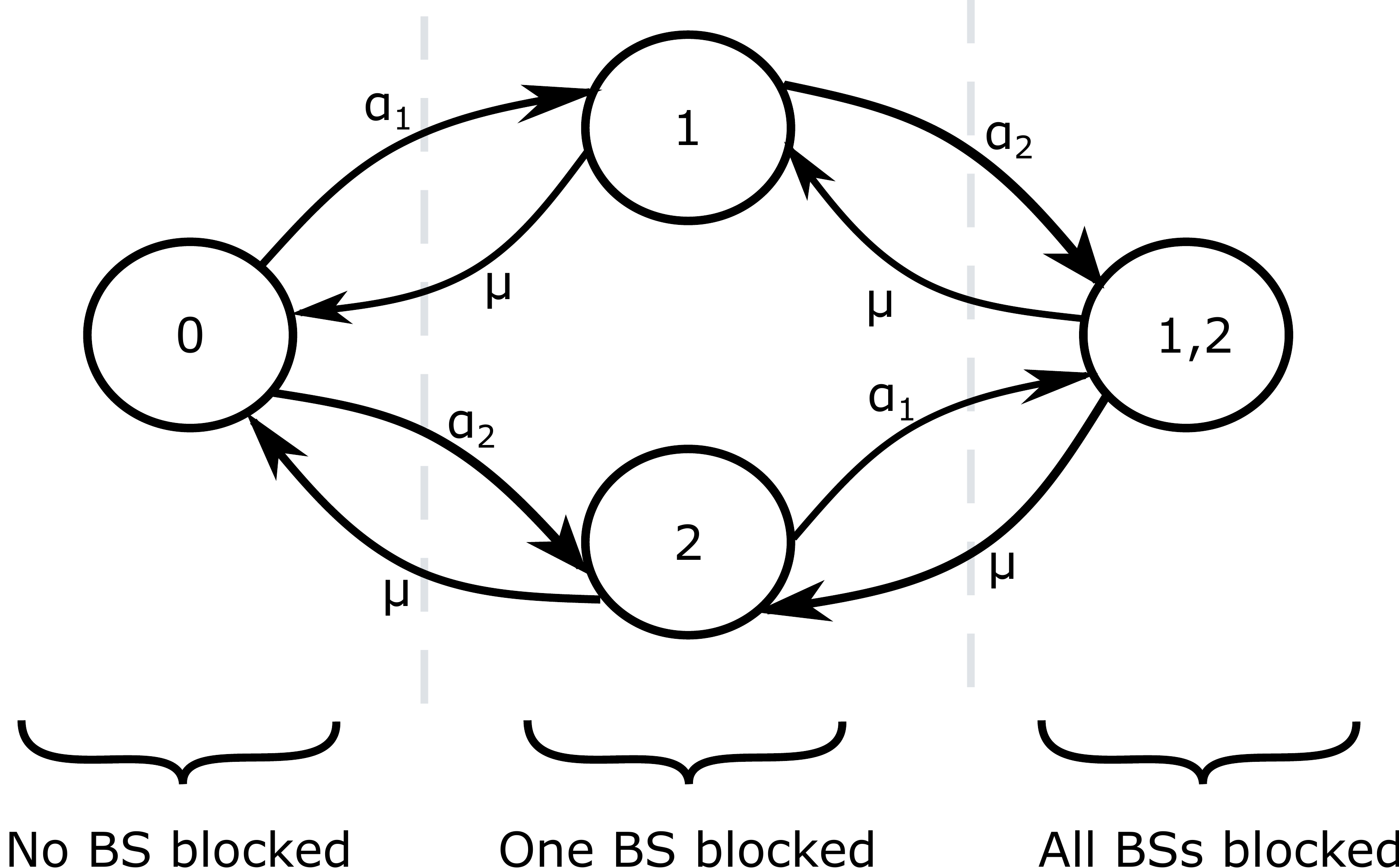}
    \caption{Markov Chain for 2 BSs}
    \label{fig:Markov2case}
\vspace{-6mm}
\end{figure}
A simplified example of 2 state Markov chain is given in Figure~\ref{fig:Markov2case}. 
 The four states in this model are described as (i) state $S_0^1$: no BS is blocked, (ii) state $S_1^1$: BS 1 is blocked, (iii) state $S_1^2$: BS 2 is blocked, and (iv) state $S_2^1$: both BS 1 and 2 are blocked. The state probabilities of the four states can be calculated as
\begin{equation}\label{eqn:Markov2case1}
\begin{split}
     P_1^1 = \frac{\alpha_1}{\mu}P_0^1;\quad\quad P_1^2 = \frac{\alpha_2}{\mu}P_0^1; \quad\quad P_{2}^1 &= \frac{\alpha_1\alpha_2}{\mu^2}P_0^1,
\end{split}
\end{equation}
where $\alpha_i$ is the blocking rate of the $i^{th}$ BS for which an expression is obtained in~(\ref{eqn:singleBS}). As the sum of probabilities of all states is equal to 1, we get the probability $P_0^1$ as,
\begin{equation}\label{eqn:Markov2case2}
\begin{split}
        P_0^1 =\frac{1}{1+\frac{\alpha_1}{\mu}+\frac{\alpha_2}{\mu}+\frac{\alpha_1\alpha_2}{\mu^2}} =\frac{1}{\left(1+\frac{\alpha_1}{\mu}\right)\left(1+\frac{\alpha_2}{\mu}\right)}.
\end{split}
\end{equation}
Finally, the Probability $P_{2}^1$ that both BSs are blocked is calculated using~(\ref{eqn:Markov2case1}) and~(\ref{eqn:Markov2case2}) as
\begin{equation}\label{eqn:p12}
    P_{1,2}=\frac{\frac{\alpha_1\alpha_2}{\mu^2}}{\left(1+\frac{\alpha_1}{\mu}\right)\left(1+\frac{\alpha_2}{\mu}\right)}.
\end{equation}

The complete analysis of blockage events is provided in the next chapter.

 \chapter{Blockage Events}\label{ch03_blockage_events}
\chaptermark{Blockage Events}

\section{Analysis of Blockage Events}

We define an indicator random variable $B$ that indicates the blockage of all available BSs in the range of UE. The blockage probability $P(B|N,\{R_i\})$ is conditioned on the number of BSs $N$ in the disc $B(o,R)$ which are not blocked by the user's body and the distances $R_i$ of BS $i = 1,\cdots,n$ from the UE. This probability is same as the state probability $P_n^1$ in  (\ref{eqn:prob_stateN})


\begin{equation}\label{eqn:AllBS}
        P(B|N,\{R_i\})=P_n^1  =\prod_{i=1}^n\frac{(C/\mu)r_i}{1+(C/\mu)r_i}.
\end{equation} 
Note that the notation $P(B|N,\{R_i\})$ is a short version of $P_{B|N,\{R_i\}}(b|n,\{r_i\})$, where the random variables are represented in capitals and their realizations in the corresponding small letters. We keep the short notation throughout the paper for simplicity.
\section{Marginal and conditional blockage probability}
We first evaluate the conditional blockage probability $P(B|N)$ by taking the average of $P(B|N,\{R_i\})$ over the distribution of $\{R_i\}$ and then find $P(B)$ by taking the average of $P(B|N)$ over the distribution of $N$ as follow
\begin{equation}\label{eqn:pBgivenN}
\begin{split}
P(B|N)=\!\!\int\!\!\!\int_{r_i}\!\! P(B|N,\{R_i\})\;f(\{R_i\}|N)\; dr_1\cdots dr_n
 \end{split}
\end{equation}
\begin{equation}\label{eqn:ep1n}
\begin{split}
   P(B)=\sum_{n=0}^{\infty} P(B|N)P_N(n).
\end{split}
\end{equation}

\noindent
\begin{theorem} \label{th1} The marginal blockage probability and the conditional blockage probability conditioned on the coverage event (\ref{eqn:coveragep}) is 

\begin{equation}\label{eqn:expblockage}
        P(B) = e^{-a p\lambda_T\pi R^2},
\end{equation} 
\begin{equation}\label{eqn:expblockagecond}
       P(B|\mathcal{C}) = \frac{e^{-a p\lambda_T\pi R^2} - e^{-p\lambda_T\pi R^2}}{1-e^{-p\lambda_T\pi R^2}},
\end{equation} 
where,
\begin{equation}\label{eqn:a}
\begin{split}
      a=\frac{2\mu}{RC}-\frac{2\mu^2}{R^2C^2} \log\left(1+\frac{RC}{\mu}\right).
   \end{split}
\end{equation} 
Note that $C$ is proportional to blocker density $\lambda_B$ shown in (\ref{eqn:C}) and $p=1-\omega/2\pi$ is defined in (\ref{eqn:p}). 
\end{theorem}
\begin{proof} 
We first derive $P(B|N)$ in (\ref{eqn:pBgivenN}) as
\begin{equation}\label{eqn:proof1-firsthalf}
\begin{split}
&P(B|N)=\int\!\!\!\int_{r_i}\!\! P(B|N,\{R_i\})\;f(\{R_i\}|N)\; dr_1\cdots dr_n\\
&=\int\!\!\!\int_{r_i}\prod_{i=1}^n \frac{(C/\mu)r_i}{1+(C/\mu)r_i} \frac{2r_i}{R^2} dr_i \\
& = \prod_{i=1}^n \int_{r=0}^{R} \frac{(C/\mu)r}{1+(C/\mu)r} \frac{2r}{R^2} dr \\
&=\left(\int_{r=0}^R\frac{(C/\mu)r}{1+(C/\mu)r} \frac{2r}{R^2}\,dr\right)^n\\
&= \left(\int_{r=0}^R\left(\frac{2r}{R^2}-\frac{2\mu}{R^2C}+\frac{2\mu}{R^2C}\frac{1}{(1+Cr/\mu)}\right)\,dr\right)^n\\
&=\left(\left(\frac{r^2}{R^2}-\frac{2\mu r}{R^2C}+\frac{2\mu^2}{R^2C^2}\log(1+Cr/\mu)\right)\bigg|_0^R\right)^n\\
&=\left(1-\frac{2\mu }{RC}+\frac{2\mu^2}{R^2C^2}\log(1+RC/\mu)\right)^n\\
&=(1-a)^n,
 \end{split}
\end{equation}
where $a$ is given in (\ref{eqn:a}).
Next, we evaluate $P(B)$ in (\ref{eqn:ep1n}) as
\begin{equation*}\label{eqn:probPBpart2Dyn}
\begin{split}
   &P(B)=\sum_{n=0}^{\infty} P(B|N)P_N(n),\\
   &=\sum_{n=0}^\infty (1-a)^n  \frac{[p\lambda_{T} \pi R^2]^n}{n!}e^{-p\lambda_{T} \pi R^2} \\
& = e^{-ap\lambda_{T} \pi R^2}\sum_{n=0}^\infty \frac{[(1-a)\lambda_{T} \pi R^2]^n}{n!}e^{-(1-a)\lambda_{T} \pi R^2}\\
&= e^{-ap\lambda_{T} \pi R^2}. \\
\end{split}
\end{equation*}

Finally, the conditional blockage probability $P(B|\mathcal{C})$ conditioned on coverage event is obtained as
\begin{equation}\label{eqn:pureBlk}
\begin{split}
P(B|\setC)& = \frac{P(B,\setC)}{P(\setC)} =\frac{\sum_{n=1}^{\infty} P(B|N)P_N(n)}{P(\setC)}\\
&=\frac{e^{-a p\lambda_T\pi R^2} - e^{-p\lambda_T\pi R^2}}{1-e^{-p\lambda_T\pi R^2}},
\end{split}
\end{equation}

This concludes the proof of Theorem~\ref{th1}.



\end{proof}
We observed from Theorem \ref{th1} that the expected probability of simultaneous blockage decreases exponentially with the BS density $\lambda_T$. Further, note that $a\in(0,1)$, where $a\rightarrow 1$ when $RC/\mu\rightarrow 0$ and $a\rightarrow 0$ when $RC/\mu\rightarrow \infty$. Since the upper bound is trivial, we only prove the lower bound. Consider the series expansion of $\log(1+RC/\mu)$ in $a$, \textit{i.e.},
\begin{equation}
\begin{split}
a&= \frac{2\mu}{RC}-\frac{2\mu^2}{R^2C^2} \left(\frac{RC}{\mu}-\frac{R^2C^2}{2\mu^2}+\frac{R^3C^3}{3\mu^3}+\cdots\right)\\
&\approx 1-\frac{2RC}{3\mu}, \quad \text{when } \frac{RC}{\mu} \text{ is small}.
\end{split}
\end{equation}

Thus, when $RC/\mu\rightarrow 0$, then $a\rightarrow 1$. 
\begin{figure}[!t]
	\centering
	\includegraphics[width=0.8\textwidth]{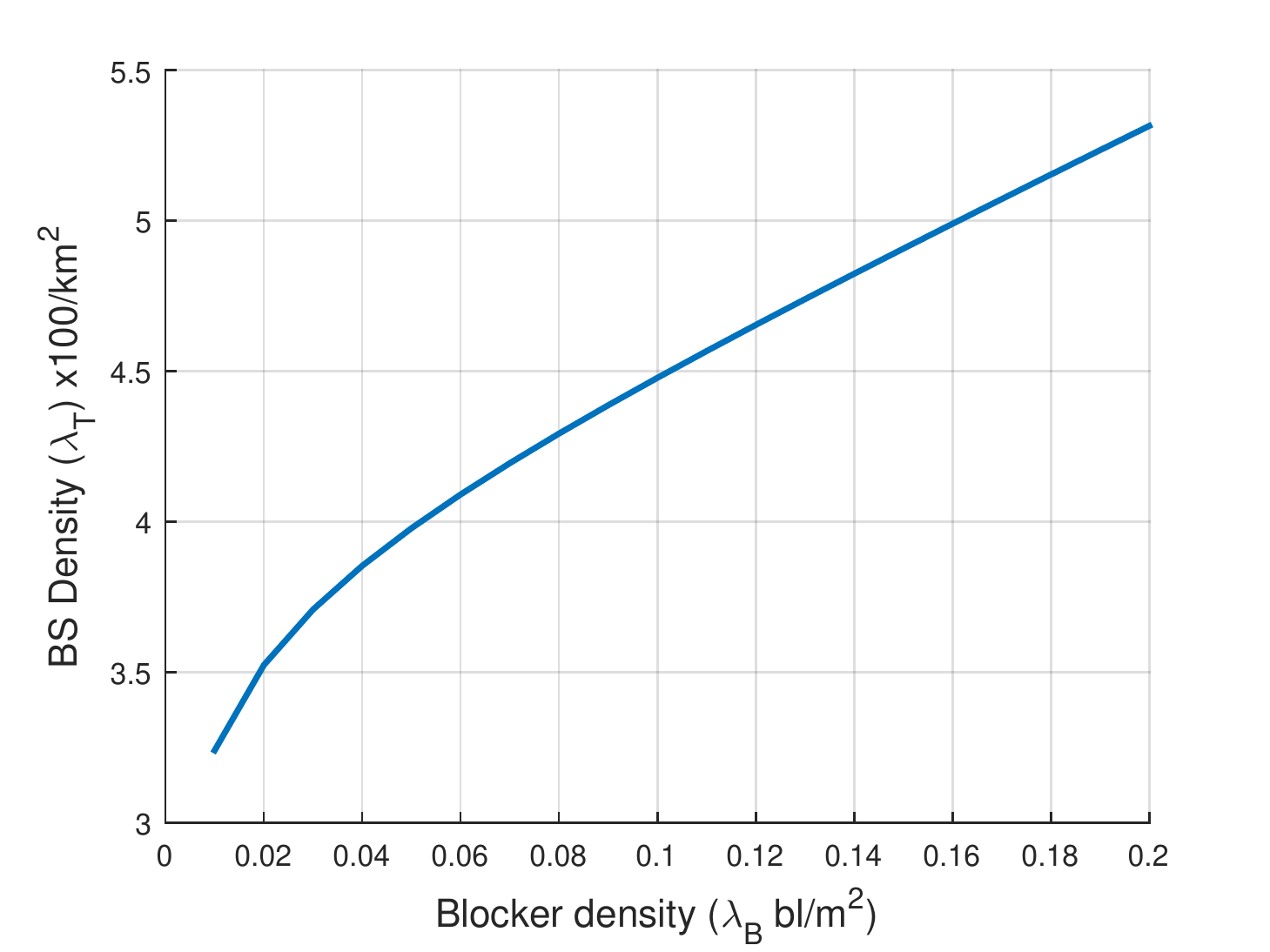}
    \captionsetup{skip=-2pt}
	\caption{BS density vs blocker density for $\bar{P}=1e-5$.}
	\label{fig:BS_bl_density}
	\vspace{-5mm}
\end{figure}

Note that for the blocker density as high as $0.1$ bl/m$^2$, and for other parameters in Table \ref{SimParams}, we have $RC/\mu = 0.35$, which shows that the approximation holds for a wide range of blocker densities.
For large BS density $\lambda_T$, the coverage probability $P(\setC)$ is approximately 1 and $P(B|\setC)\approx P(B)$. 
In order to have the blockage probability $P(B)$ less than a threshold $\bar{P}$
\begin{equation}
P(B)=e^{-ap\lambda_T\pi R^2}\leq\bar{P},
\end{equation}
the required BS density follows
\begin{equation} \label{eqn:approx_lamT}
\lambda_T\geq\frac{-\log(\bar{P})}{ap\pi R^2}\approx \frac{-\log(\bar{P})(1+\frac{2RC}{3\mu})}{p\pi R^2},
\end{equation}
where $C$ is proportional to the blocker density $\lambda_B$ in (\ref{eqn:C}). Thus, the approximation holds for smaller $\lambda_B$. The result (\ref{eqn:approx_lamT}) shows that the BS density approximately scales linearly with the blocker density and is plotted in Figure \ref{fig:BS_bl_density} for $\bar{P}=1e-5$ and other parameters in Table \ref{SimParams}.




\section{Expected Blockage Frequency}
From the Markov model in Figure~\ref{fig:MarkovNcase}, we know that the total arrival rate of blockers in the state when all BSs get simultaneously blocked is same as the total departure rate from that state in the equilibrium. Therefore, the frequency/rate of simultaneous blockage of all $N$ BSs is:

\begin{equation}\label{eqn:AllBSfreq}
\begin{split}
    \zeta_B = n\mu P(B|N,\{R_i\})=n\mu \prod_{i=1}^n\frac{(C/\mu)r_i}{1+(C/\mu)r_i},
\end{split}
\end{equation}

Thus, the expected rate of blockage is obtained where the expectation is taken over the joint distribution of $N$ and $\{R_i\}$.
\begin{equation}\label{eqn:blockageDurEQn1}
\E[\zeta_B|N] = \!\!\int\!\!\!\int_{r_i}\!\! \zeta_B\; f(\{R_i\}|N)\;dr_1\cdots dr_n,
\end{equation}
\begin{equation}\label{ep1n}
        \mathbb{E} \left[\zeta_B\right] = \sum_{n=0}^\infty \E[\zeta_B|N]P_N(n).
\end{equation}

\begin{theorem} \label{th2} The expected frequency of simultaneous blockage of all BSs in the disc of radius $R$ around UE is 
\begin{equation}\label{eqn:blkfrq}
        \mathbb{E}\left[\zeta_B\right] = \mu (1-a)p \lambda_T\pi R^2 e^{-ap\lambda_T\pi R^2},
\end{equation}       and the expected frequency conditioned on the coverage event (\ref{eqn:C}) is \begin{equation}\label{eqn:blkfreq_cond}
\begin{split}
\MoveEqLeft \mathbb{E}\left[\zeta_B|\setC\right] = \frac{\mu (1-a)p\lambda_T\pi R^2e^{-ap\lambda_T\pi R^2}}{{1-e^{-p\lambda_T\pi R^2}}},
\end{split}
\end{equation}
where $a$ is defined in (\ref{eqn:a}).
\end{theorem}
\begin{proof}
We first evaluate $\E[\zeta_B|N]$ given in (\ref{eqn:blockageDurEQn1}) \textit{i.e.}, 
\begin{equation}
\begin{split}
\E[\zeta_B|N] &= n\mu \!\!\int\!\!\!\int_{r_i} \prod_{i=1}^n \frac{(C/\mu)r_i}{1+(C/\mu)r_i} \frac{2r_i}{R^2} dr_i \\
& = n\mu (1-a)^n,
\end{split} 
\end{equation}
and then we evaluate $\E[\zeta_B]$ given in (\ref{ep1n}), \textit{i.e.},
\begin{equation}
\begin{split}
 &\mathbb{E} \left[\zeta_B\right]  =  \sum_{n=0}^\infty  n \mu(1-a)^n \frac{[p\lambda_{T} \pi R^2]^n}{n!}e^{-p\lambda_{T} \pi R^2} \\
 &= \mu(1-a)p\lambda_{T} \pi R^2e^{-ap\lambda_{T} \pi R^2}\times\\
 &\qquad\quad\sum_{n=0}^\infty  \frac{[(1-a)p\lambda_{T} \pi R^2]^{(n-1)}}{(n-1)!}e^{-(1-a)p\lambda_{T} \pi R^2} \\
 &= \mu(1-a)p\lambda_{T} \pi R^2e^{-ap\lambda_{T} \pi R^2}.\\
\end{split}
\end{equation}
Finally, the expected frequency of blockage conditioned on the coverage events (\ref{eqn:coveragep}) is given by

\begin{equation}
\begin{split}
\MoveEqLeft \mathbb{E} \left[\zeta_B|\setC\right] = \frac{\sum_{n=1}^\infty \E[\zeta|N] P_N(n)}{P(\setC)} =\frac{\sum_{n=0}^\infty \E[\zeta|N] P_N(n)}{P(\setC)} \\
 &= \frac{\mu(1- a)p\lambda_T\pi R^2e^{-ap\lambda_T\pi R^2}}{{1-e^{-p\lambda_T\pi R^2}}}.
\end{split}
\end{equation} 
This concludes the proof of Theorem~\ref{th2} 

\end{proof}


  

\section{Expected Blockage Duration}
Recall that the duration of the blockage of a single BS-UE link is an exponential random variable $T_i \sim \exp(\mu)$, \textit{i.e.},
\begin{equation}
f_{T_i}(t_i) = \mu e^{-\mu t_i},\quad \text{for} \;\;i=1:n.
\end{equation}
We show that the duration of the blockage of all $n$ BSs follows an exponential distribution with mean $1/n\mu$.   
Consider a time instant when all $n$ BSs are blocked; the residual duration of the blockage period of the $i^{\text{th}}$ BS-UE link follows the same distribution as $f_{T_i}(t_i)$ because of the memoryless property of the exponential distribution.
Therefore, the duration of the period of simultaneous blockage of all $n$ BSs is a random variable $T_B=\min\{T_1,T_2,\cdots,T_n\}$. Note that $T_B$ follows the distribution 
$T_B \sim \exp(n\mu)$, conditioned on the number of BSs $N=n$. We can write the expected blockage duration as 
\begin{equation}\label{eqn:exp_TBgivenN}
\E[T_B|N] = \frac{1}{n\mu}.
\end{equation}

\noindent
\begin{theorem} The expected blockage duration of the period of the simultaneous blockage of all the BSs in $B(o,R)$ conditioned on the coverage event $\setC$ in (\ref{eqn:coveragep}) is obtained as
\begin{equation}\label{eqn:blockageDurationExp}
        \mathbb{E}\left[T_B|\setC\right] = \frac{e^{-p\lambda_T\pi R^2}}{\mu\left(1-e^{-p\lambda_T\pi R^2}\right)}\text{Ei}\left[p\lambda_T\pi R^2\right].
\end{equation}        
where, $\text{E}\text{i}\left[p\lambda_T\pi R^2\right]
 =  \int_{0}^{p\lambda_T\pi R^2} \frac{e^x-1}{x} dx$ 
= $\sum_{n=1}^\infty\frac{[p\lambda_T\pi R^2]^n}{nn!}$.
\end{theorem}
\begin{proof}
Using the results from (\ref{eqn:exp_TBgivenN}), we find the expected blockage duration $\E[T_B|\setC]$ conditioned on the coverage event $\setC$ defined in (\ref{eqn:coveragep}) as follow
\begin{equation}
\begin{split}
 \mathbb{E} \left[T_B|\setC\right] &= \frac{\sum_{n=1}^\infty \frac{1}{n\mu} P_N(n)}{P(\setC)} \\
&= \frac{\sum_{n=1}^\infty \frac{1}{n\mu} \frac{[p\lambda_{T} \pi R^2]^n}{n!}e^{-p\lambda_{T} \pi R^2}}{1-e^{-p\lambda_T\pi R^2}} \\
&= \frac{e^{-p\lambda_{T} \pi R^2}}{\mu\left(1-e^{-p\lambda_T\pi R^2}\right)}\sum_{n=1}^\infty \frac{[p\lambda_{T} \pi R^2]^n}{n n!}. 
\end{split}
\end{equation}

Let us consider the series expansion of $e^x$.
\begin{equation}
\begin{split}
\MoveEqLeft e^x = 1+x+\frac{x^2}{2!}+\frac{x^3}{3!}+\frac{x^4}{4!}+\frac{x^5}{5!}+\cdots \qquad \qquad \qquad \\
\MoveEqLeft e^x = 1+\sum_{n=1}^\infty \frac{x^n}{n!} \implies e^x-1 = \sum_{n=1}^\infty \frac{x^n}{n!} \\ \MoveEqLeft \implies \frac{e^x-1}{x} = \sum_{n=1}^\infty \frac{x^{n-1}}{n!}.\\
\end{split}
\end{equation}
Integrating both side, we have
\begin{equation}
\begin{split}
\MoveEqLeft \int_{0}^{\lambda_T\pi R^2} \frac{e^x-1}{x} dx = \sum_{n=1}^\infty \int_{0}^{\lambda_T\pi R^2} \frac{x^{n-1}}{n!} dx\\
\MoveEqLeft \text{E}\text{i}\left[\lambda_T\pi R^2\right] = \int_{0}^{\lambda_T\pi R^2} \frac{e^x-1}{x} dx = \sum_{n=1}^\infty \frac{[\lambda_{T} \pi R^2]^n}{n n!}.
\end{split}
\end{equation}
Hence,
\begin{equation*}
\mathbb{E}\left[T_B|\setC\right] = \frac{e^{-\lambda_T\pi R^2}}{\mu\left(1-e^{-\lambda_T\pi R^2}\right)}\text{E}\text{i}\left[\lambda_T\pi R^2\right].
\end{equation*}

\end{proof}

\noindent
\begin{lemma}
$\text{E}\text{i}\left[\lambda_T\pi R^2\right]$ converges.
\end{lemma} 
\begin{proof}
 We can use Cauchy ratio test to show that the series $\sum_{n=1}^\infty\frac{[\lambda_T\pi R^2]^n}{nn!}$ is convergent. Consider 
$L = \lim_{n\rightarrow\infty}\frac{[\lambda_T\pi R^2]^{n+1}/((n+1)(n+1)!)}{[\lambda_T\pi R^2]^n/(nn!)} = \lim_{n\rightarrow\infty}\frac{[\lambda_T\pi R^2]n}{(n+1)^2}=0$. Hence, the series converges.
\end{proof}
An approximation of blockage duration can be obtained for a high BS density as follow
\begin{equation}
\E[T_B|\setC]\approx \frac{1}{\mu p\lambda_T\pi R^2}+ \frac{1}{(\mu p\lambda_T\pi R^2)^2}.
\end{equation}
This approximation is justified as follow
The expectation of a function $f(n) = 1/n$ can be approximated using Taylor series as
\begin{equation}
\begin{split}
\E[f(n)] &= \E(f(\mu_n+(x-\mu_n)))\\
&=\E[f(\mu_n)+f'(\mu_n)(n-\mu_n)+\frac{1}{2}''(\mu_n)(x-\mu_n)^2]\\
&\approx f(\mu_n)+\frac{1}{2}f''(\mu_n)\sigma_n^2\\
&=\frac{1}{\mu_n}+\frac{\sigma_n^2}{\mu_n^3},
\end{split}
\end{equation}
where $\mu_n$ and $\sigma_n^2$ are the mean and variance of Poisson random variable $N$ given in (\ref{eqn:PN}). We get the required expression by using $\mu_n = p\lambda_T\pi R^2$ and $\sigma_n^2 = p\lambda_T\pi R^2$.

 \chapter{Numerical Evaluation}\label{ch04_results}
\chaptermark{Numerical Evaluation}


\section{Simulation Setup}
\label{sec:numResults}
This section compares our analytical results with MATLAB simulation\footnote{Our simulator MATLAB code is available at github.com/ishjain/mmWave.} 
where the movement of blockers is generated using the random waypoint mobility model~\cite{randomWayPoint,randomWayPointSim}. For the simulation, we consider a rectangular box of $200 \ \text{m}\times 200 \ \text{m}$ and blockers are located uniformly in this area.
Our area of interest is the disc $B(o,R)$ of radius $R=100 $m, which perfectly fits in the considered rectangular area. 
The blockers chose a direction randomly, and move in that direction for a time-duration of $t\sim \text{Unif}[0,60] \ \text{sec}$. To maintain the density of blockers in the rectangular region, we consider that once a blocker reaches the edge of the rectangle, they get reflected. 

We used the Mathwork code
for this purpose. The simulation runs for an hour. We note the time instant when the blocker crosses a BS-UE link and generate a blockage duration through a realization of an exponential distribution with mean $\mu=2$. Further, we collect the time-series of alternate blocked/unblocked intervals for all the BS-UE links and take their intersection to obtain a time-series that represent the events of blockage of all available BSs. The blockage probability, frequency, and duration can be obtained from this time-series. Finally, we repeat the procedure for 10,000 iterations and report the average results. Rest of the simulation parameters are presented in Table~\ref{SimParams}.   

\begin{table}[H]
\vspace{2mm}
\caption{Simulation parameters}
\label{SimParams}
\centering
\begin{tabular}{|c|c|}\hline
	\textbf{Parameters} & \textbf{Values} \\\hline 
    Radius  $R$ &100 m \\
    Velocity of blockers $V$&1 m/s\\
    Height of Blockers $h_B$ & 1.8 m\\
    Height of UE $h_R$ & 1.4 m\\
    Height of APs $h_T$ & 5 m\\
    Expected blockage duration $1/\mu$ &1/2 s\\
    Self-blockage angle $\omega$ & 60$^\text{o}$ \\
   
\hline
\end{tabular}
\vspace{-6mm}
\end{table}
\section{Main Results}

\begin{figure}[!t]
	\centering
	\includegraphics[width=0.7\textwidth]{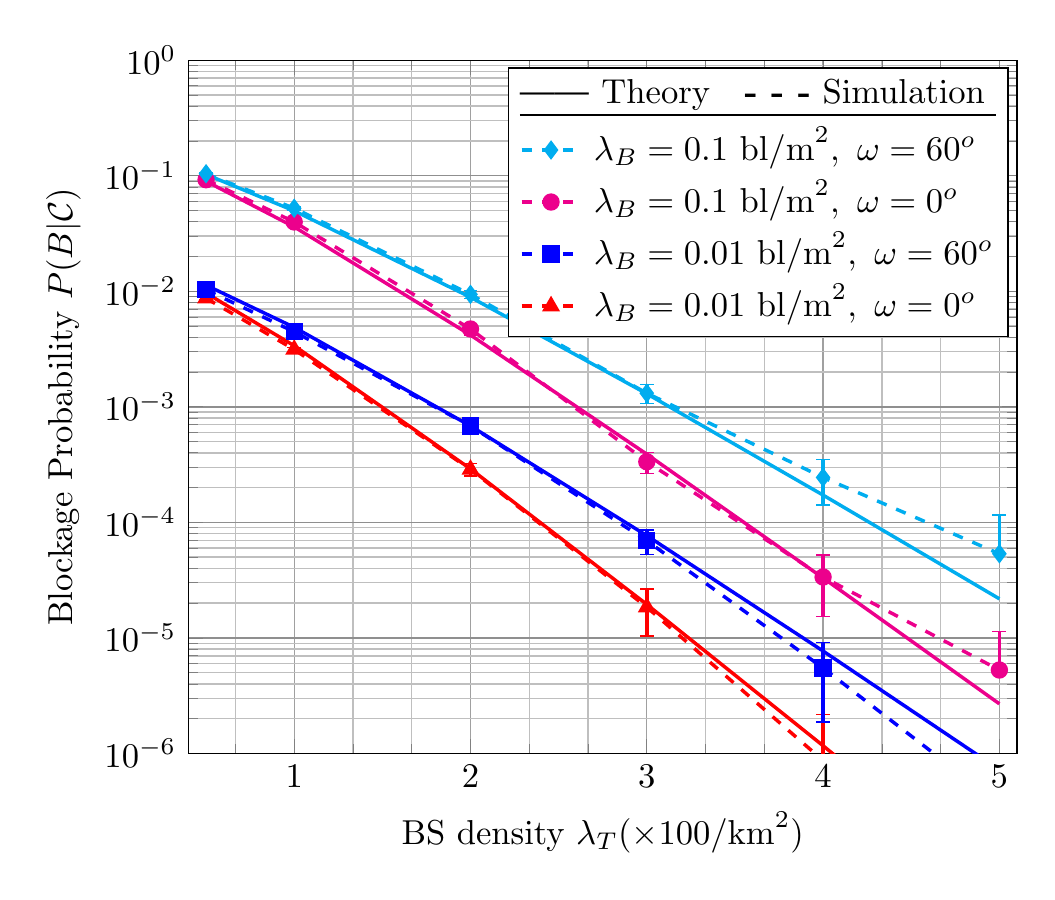}
	\caption{Conditional blockage probability }
	\label{fig:condBlProb}
\end{figure}

\begin{figure}[!t]
	\centering
	\includegraphics[width=0.7\textwidth]{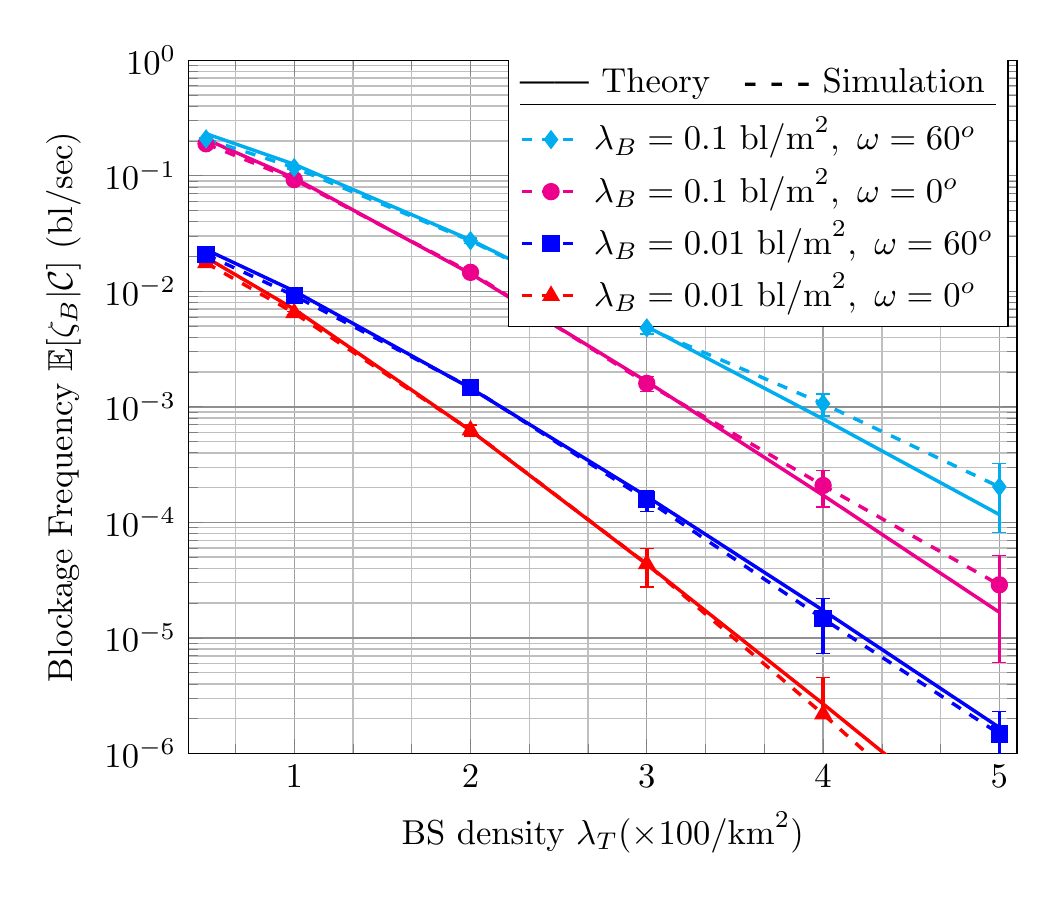}
	\caption{Conditional blockage frequency}
	\label{fig:condBlfrac}
\end{figure}

\begin{figure}[!t]
	\centering
	\includegraphics[width=0.7\textwidth]{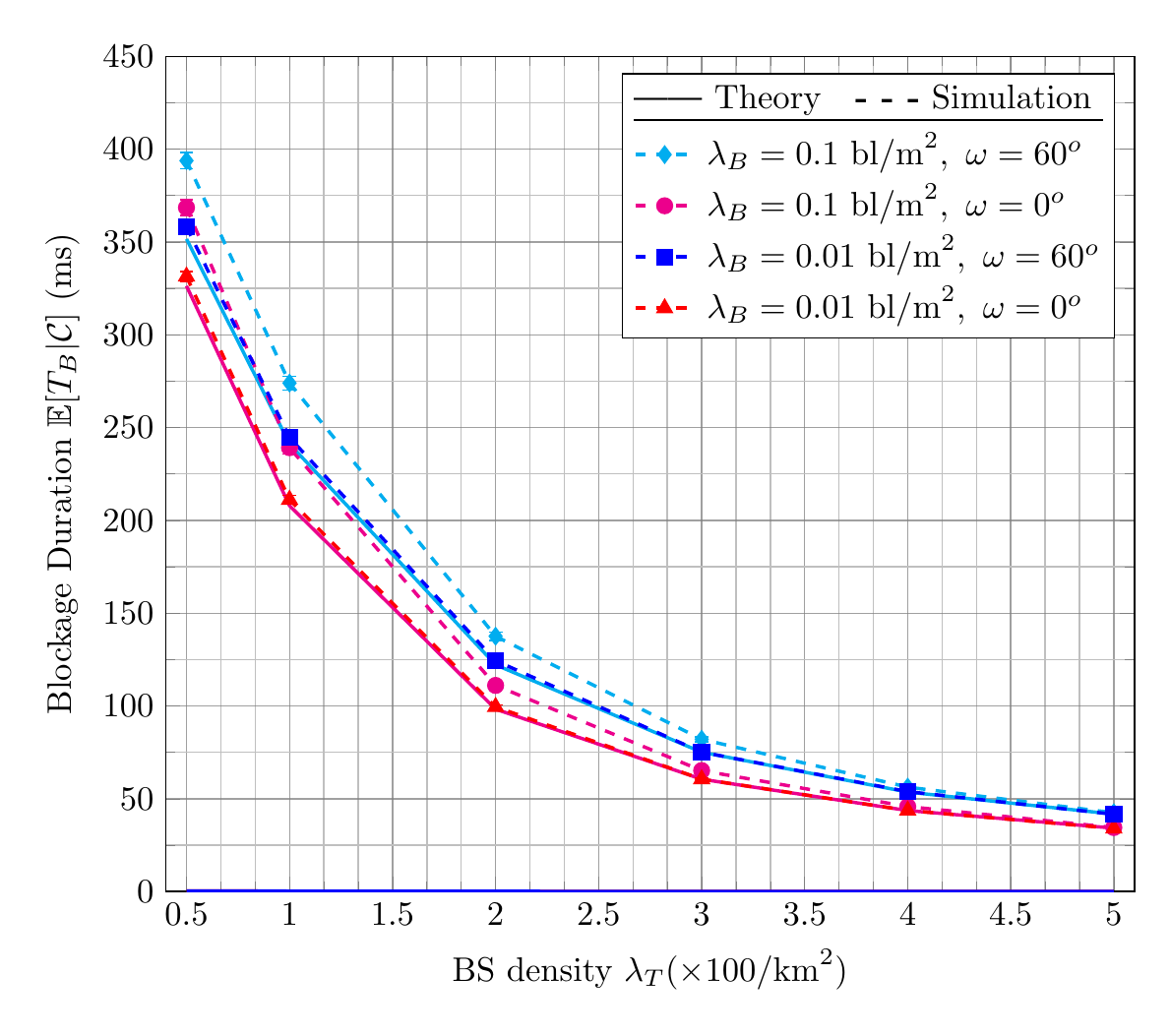}
	\caption{Conditional blockage duration }
	\label{fig:condBldur} 
\end{figure}

We present the comparison between our analytical and simulation results with the joint impact of the dynamic and self-blockages. 

We consider two values of mobile blocker density, $0.01$ and $0.1$ bl/m$^2$, and two values of the self-blocking angle $\omega$ ($0$ and $\pi/3$) for our study. Figures~\ref{fig:condBlProb},~\ref{fig:condBlfrac}, and~\ref{fig:condBldur} present the statistics of blockages when the UE has at least one serving BS, \textit{i.e.}, the UE is always in the coverage area of at least one BS. From Figure~\ref{fig:condBlProb} and Figure~\ref{fig:condBlfrac}, we can observe that the blockage probability and the expected blockage frequency decrease exponentially with BS density. From the point of view of interactive applications such as AR/VR, video conferencing, online gaming, and others, this means that a higher BS density can potentially decrease interruptions in the data transmission. For example, for a blocker density of 0.1 bl/m$^2$, a BS density of $100$/km$^2$ can decrease the interruptions to once in ten seconds, $200$/km$^2$ can decrease them to once in 100 seconds, and $300$/km$^2$ 
decrease them to once in 1000 seconds. Reducing the frequency of interruptions is particularly crucial for AR/VR applications, therefore from this perspective a density of 200-300/km$^2$ may be required. This corresponds to about 6 to 9 BS, respectively,  within the range of each UE. From Figure~\ref{fig:condBldur}, we can observe that caching of $100$ ms worth of data is required for a BS density $200$/km$^2$ to have uninterrupted services. For AR and tactile applications, caching is not a solution and a delay of 100 ms may be an unacceptable delay. Switching to microwave networks such as 4G during blockage events may be an alternative solution instead of deploying a high  BS density, but then this may need careful network planning so as to not overload the 4G network. The amount of required cached data decreases with increasing BS density. A BS density of $300$/km$^2$ and $500$/km$^2$ can bring down the required cached data to $60$ ms and $40$ ms, respectively. This may be acceptable for AR/VR applications if these freezes are infrequent. Thus, the cellular architecture needs to consider the optimal amount of cached data and the optimal BS density needed to mitigate the effect of these occasional high blockage durations to satisfy QoS requirements for AR/VR application without creating nausea. A tentative conclusion is that perhaps a minimum acceptable density of 300/km$^2$ (which corresponds to about 9 BS within range of each UE) is needed to keep interruptions lasting about 60 ms to once every 1000 seconds.    

We also observe that both simulation and analytical results are approximately the same for a low blocker density of $0.01$ bl/m$^2$. 
From Figure~\ref{fig:condBldur}, we observe our analytical result deviates from the simulation result for lower values of BS densities. However, the percentage error ($\sim10-15\%$) is not significant.  Thus, our approximation in Lemma 2 is validated.



\begin{figure}[!t]
	\centering
	\includegraphics[width=0.7\textwidth]{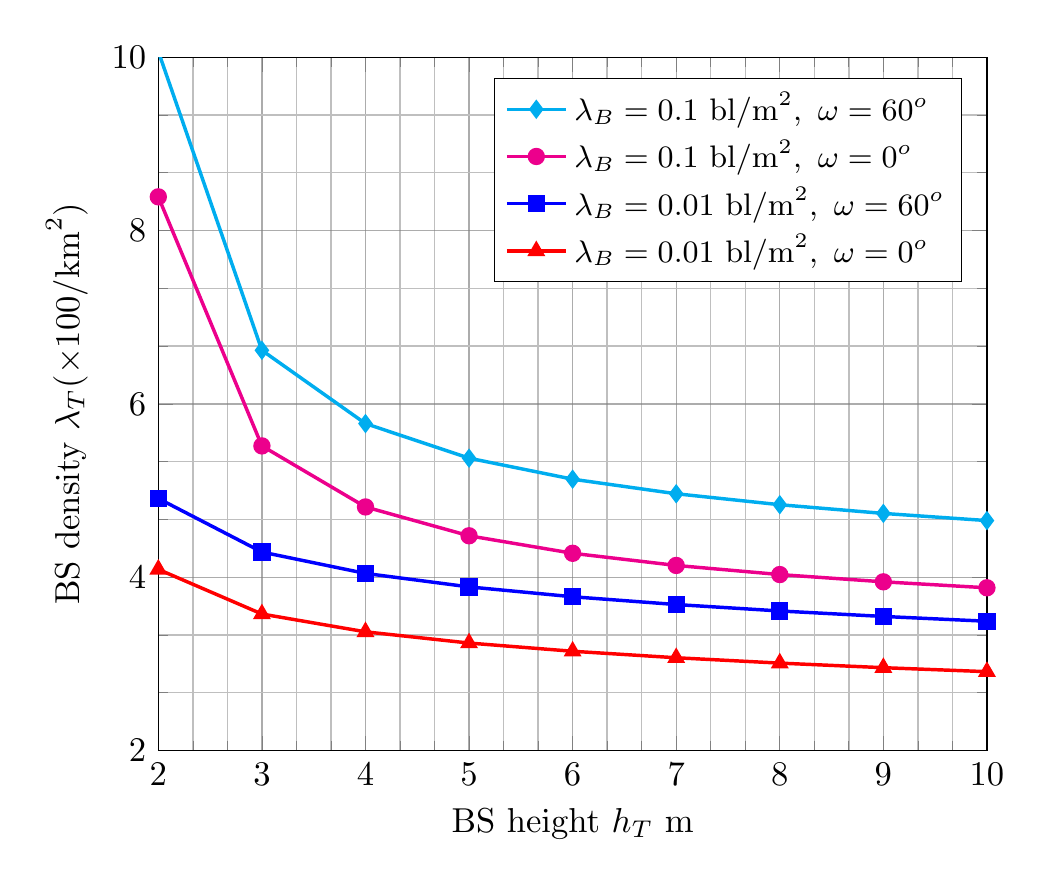}
	\caption{The trade-off between BS height and density for fixed blockage probability $P(B|\setC)=1e-7$. }
	\label{fig:heightvsdensity}
	\vspace{-6mm}
\end{figure}

\section{Case Study}
\subsection{Required minimum BS density}
5G-PPP has issued requirements for 5G use cases~\cite{mohr20165g} with  service reliability $\geq 99.999\%$ for specific mission-critical services.   
From Figure \ref{fig:condBlProb}, we can infer that the minimum BS density required for a maximum blockage probability $P(B| \setC) = 1e-5$ is 400 BS/km$^2$ for a blocker density of 0.01 bl/m$^2$ and self-blockage angle of $60^o$. For a higher blocker density, the required BS density increases linearly.
Note that this again imposes a higher BS density than would be necessary from most models based solely on capacity needs (roughly 100 BS/km$^2$~\cite{CellularCap-Rap}).

\subsection{BS density-height trade-off analysis}
The BS height vs. density trade-off is shown in Figure \ref{fig:heightvsdensity}. Note, for example, that doubling the height of the BS from 4m to 8 m reduces the BS station density requirement by approximately 20\% for blocker density $\lambda_B=0.1$  bl/m$^2$ and self-blockage angle $\omega = 60^{o}$.
The optimal BS height and density can be obtained by performing a cost analysis.

 \chapter{Conclusions and Future Work}\label{ch.conc}

\section{Conclusion}\label{sec:conc}
In this thesis, we analyzed the blockage problem in mmWave cellular networks. Specifically, we considered an open park-like scenario with dynamic blockages due to mobile humans and vehicles collectively called the mobile blockers. The blockage rate which is defined as the rate of blockage of the BS-UE LOS link by the mobile blockers is evaluated as a function of blocker density, height, and velocity.
We also considered self-blockage due to user's own body. The blockage process of a single BS-UE link is modeled as an alternating renewal process with exponentially distributed intervals of blocked and unblocked periods. We extend the blocking scenario to consider multiple BSs using stochastic geometry and a Markov chain model. In particular, we consider that the UE can instantaneously switch between the BSs in case the currently serving BS gets blocked. In this setting, the blockage event occurs when all BSs in the range of UE are simultaneously blocked. We derived the closed-form expressions for blockage probability and blockage frequency as a function of the density and height of the BS and blockers. We also evaluated an approximate expression of the blockage duration. Finally, we verified our theoretical model with MATLAB simulations considering a random waypoint mobility model of the blockers. We get the following insights from our blockage analysis

\begin{itemize}
\item The minimum density of BS required to bound the blockage probability below $1e-5$ for the blocker density of $0.01 $bl/m$^2$ and self-blockage angle $\omega=60^o$ is 400 BS/km$^2$ (effective cell size $25$m). This requirement is much higher than that obtained from capacity constraints alone.
\item The blockage duration at high BS density saturates to around 40 ms which is higher than that required for AR/VR applications.
\item The BS density can be reduced by increasing the BS height. The increase in height from 4 m to 8 m can reduce the BS density by 20\%. Further increase in height may not lead to a significant reduction in density.
\end{itemize}

\section{Future Work}\label{sec:future}
The following extensions are planned for future work
\begin{itemize}
\item Generalized blockage model: We can add a simple model for static blockage in our analysis of dynamic and self-blockage. 
\item Data rate analysis: The data rates of a typical user can be evaluated using the generalized blockage model. We are interested in evaluating whether 5G mmWave is capacity limited or blockage limited.
\item Fallback to 4G LTE: We plan to explore the potential solution to blockages as switching to 4G LTE. Whether 4G would be able to handle the huge intermittent 5G traffic.
\item Deterministic networks: We have considered a random deployment of BSs in our analysis. However, in most cases, the deployments of BSs are based on a deterministic hexagonal grid. Therefore, a blockage model for the deterministic networks is more practical.
\item Backhoul capacity analysis: With the UEs switching between BSs in case of blockage events, the backhoul capacity requirements for the BS may have high fluctuations. It is interesting to study that random process.
\end{itemize}


\appendix 



\bibliographystyle{IEEEtran}
\bibliography{references}



\end{document}